# Generative AI Adoption in Postsecondary Education, AI Hype, and ChatGPT's Launch

Isabel Pedersen
Faculty of Social Science and Humanities
Ontario Tech University

**Correspondence**:
Isabel Pedersen
Faculty of Social Science and Humanities
Ontario Tech University
Email: isabel.pedersen [at] ontariotechu.ca

**Abstract**

The rapid integration of generative artificial intelligence (AI) into postsecondary education and many other sectors resulted in a global reckoning with this new technology. This paper contributes to the study of the multifaceted influence of generative AI, with a particular focus on OpenAI's ChatGPT within academic settings during the first six months after the release in three specific ways. First, it scrutinizes the rise of ChatGPT as a transformative event construed through a study of mainstream discourses exhibiting AI hype. Second, it discusses the perceived implications of generative AI for writing, teaching, and learning through the lens of critical discourse analysis and critical AI studies. Third, it encourages the necessity for best practices in the adoption of generative AI technologies in education.

**Keywords**: AI and education, educational technologies, AI hype, ChatGPT, critical discourse analysis

**Introduction**

The launch of OpenAI's ChatGPT chatbot in November 2022 made generative artificial intelligence, "Generative AI," a mainstream phenomenon.[1] Simultaneously, the sudden prominence of generative AI sent ripples through postsecondary educational landscapes. One poll reported that "30% of college students used ChatGPT to do their schoolwork" in 2022 (Intelligent, 2023, para. 2). One might characterize the launch as a shocking societal turn. ChatGPT gained one million users in days following its launch (Thormundsson, 2023). One year later, it has 100 million active users (Malik, 2023), with more than 92% of Fortune 500 companies using the platform in multiple industries[2] (Field, 2023). Companies are progressing rapidly toward adopting AI technologies as a foundational Internet infrastructure, suggesting that AI chatbot interactions will replace traditional search engine interfaces. While artificial intelligence companies continue to develop and release conversational AI chatbots through a perpetual race to gain market share, ChatGPT's emergence in 2022 is unique; it is the first significant generative AI platform to be adopted en masse. However, the dramatic hype surrounding its release also unhinged postsecondary education professionals from a measured approach to its adoption. Postsecondary usage of it in Canadian educational spheres has been described as "ad hoc, uneven, unequal, experimental, and largely guided by individual faculty" (Veletsianos 2023b, p. 2). As an article in the *Times Higher Education* explains, "opinion pieces, blogs and social media feeds are full of questions with no clear answers, yet educators are having to grapple with AI concerns in their classes" (Chami, 2023, para. 1). The discourse surrounding ChatGPT—the first generative AI product to be hyped across public media outlets—not only helped lead to student adoption of generative AI, but it also marred an appropriate, value-based conversation about it in terms of postsecondary education.

This article argues that this discursive landscape has affected educators in all professional roles, asking them to adopt and adapt to technology that was, to a certain extent, imposed on them, obfuscating reasonable practices and the establishment of policies concerning the use of generative AI as an educational technology. One effect of AI hype is that students, teachers, administrators, teaching and learning specialists, policymakers, and legislators adapt to a computing paradigm that is encroaching on society without any guardrails from the very people who invented these technologies. Hyped discourses not only sensationalize, but they also instill values that are difficult to ameliorate. They produce "values via discourse" that encompass "their legitimation and transformation into taken-for-granted assumptions in the public domain" (Omrow, 2018, p. 15). Through emotional appeals to the technology's superiority and threats of inevitable further deployment, students adopted generative AI during a heavily hyped moment in generative AI's emergence.

This article discusses an ongoing case study that concentrates on the popular adoption of generative AI usage to better understand how it is affecting adaptation in postsecondary spheres.[3] It analyzes how generative AI, while assisting in creative and academic endeavors, also challenges fundamental notions of writing, learning, and intellectual development. It also finds a rhetoric in legacy news media discourses, indicating significant hype. This article describes this finding through a critical discourse analysis on two publications considered to be authoritative and independent, a Canadian newspaper of record, *The Globe and Mail* (readership 2.8 million), and *The Economist*, a British print and online publication (readership 6.5 million across 205 countries; Economist Impact, 2022; The Globe and Mail Newspaper Media Kit, 2023; The Economist Media Kit, 2023;). Journalistic news circulating on multiple platforms is one means through which postsecondary stakeholders first gained access to information about ChatGPT. However, this study also revealed a crisis of ethos; the credibility of experts was significantly challenged. AI scientists, professors, inventors, and CEOs of large international corporations expressed alarm and mistrust of their own AI products/inventions/research, calling into question the potential for AI to eventually harm society. Figure 1 displays four sensationalized events publicized in the news relating to AI between November 2022 and June 2023. While these events themselves are not part of the discussion, they are part of the discursive ground for this case study.

This article includes a short, selective literature review with relevant definitions. The methodology section describes how the critical discourse analysis was conducted on the news sources. Two parts analyze ChatGPT as a cultural phenomenon. The first part discusses ChatGPT's emergence as an appealing event in postsecondary education, with an emphasis on generative AI as a tool for students and instructors. It explores a paradigm shift in writing and composition that occurs when students adopt generative AI. The second part discusses ways that the launch served as a threat to traditional postsecondary values. It explores AI hype in the discourses of the legacy news sources and finds four themes to substantiate the thesis. Following that, the discussion section covers how the themes function intertextually to form AI hype. Finally, the conclusion summarizes the paper and offers a brief suggestion for next steps.

## Literature Review

On November 30, 2022, OpenAI released a free product called ChatGPT, a *generative AI chatbot*, which can be defined as a category of AI tool that uses both Natural Language Processing (NLP) technology to *interpret* human language and Natural Language Generation (NLG) technology to *produce* human language. A Generative Pretrained Transformer (GPT) is a type of Large Language Model (LLM). An LLM is

> trained on large amounts of text data for NLP tasks and contains a significant number of parameters, usually exceeding 100 million [META's Llama LLM, for example, is 65 billion parameters in size]. They [GPTs] facilitate the processing and generation of natural language text for diverse tasks. (Goyal et al., 2023, para. 8).

GPTs use machine learning and specifically, deep learning, both branches of artificial intelligence (AI) to generate text that appears human-like (hence the term *chatbot*). GPTs are called "generative" because

> they can generate new text based on the input they receive, 'pretrained' because they are trained on a large corpus of text data before being fine-tuned for specific tasks, and 'transformers' because they use a transformer based neural network architecture to process input text and generate output text [or other modes]. (Larsen & Narayan, 2023, para. 3).

While there are many generative AI chatbots on the market now, including Microsoft's Copilot, Anthropic's Claude, and Google's Gemini, this article analyzes news articles surrounding ChatGPT as the first publicly celebrated example of generative AI technology, and the one that was widely available to postsecondary students and professors between November 2022 and May 2023.

The coming chaotic era instigated due to the release of large language models was predicted by many due to earlier releases of GPTs. As Floridi and Chiriatti (2020) wrote with certainty, "GPT-3 is an extraordinary piece of technology, but as intelligent, conscious, smart, aware, perceptive, insightful, sensitive and sensible (etc.) as an old typewriter" (p. 690). They predicted the immense impact it would have but also the incredible inability for this technology to produce factual or credible outputs. Others have written about the evolving emergence of AI writers, automated writing, robo-journalism, and digital assistants (Duin & Pedersen 2021, 2023; McKee & Porter 2020). Heidi McKee and James Porter explain that

> Some of the machine writing systems that we have looked at, particularly GPT-2 and now GPT-3, work arhetorically, focusing on the text and textual production, assuming that meaning lies in the produced text. They are basically an updated version of combinatorics. The textual product stands in for the process, and when that happens rhetoric is excluded, to the detriment of effective communication. (2021, p. 48)

Automated writing is alluring. GPTs produce text but forgo the writing process, leading to the near instantaneous results students get when they ask a chatbot "to write" something. Heidi McKee and James Porter discuss precursors to ChatGPT; they argue that GPT-2 and GPT-3 are problematic writing technologies hence their claim they are *arhetorical*. GPTs produce text through predictive algorithms and do not use or even understand traditional rules of composition that people use when they choose to *compose* writing, taught to students for centuries. GPTs do not interpret social contexts in the same way as people. In traditional theories of writing composition, the social and consequential functions of texts in contexts govern how people perceive writing and exchange meaning. Text functions in a social "relationship" between a human writer and audience for a purpose (rhetorical motive). By excluding the rhetorical aspect, automated writing is reduced to a product, to the "detriment of effective communication" (McKee & Porter, 2021, p. 48).

AI hype experienced in postsecondary spheres has been discussed by education studies researchers and practitioners (Akbari, 2024; Chami, 2023; Paulson, 2024;). AI hype is a rhetorical term that can be used to characterize sensationalized or hyperbolic descriptions or representations of artificial intelligence in discourses. On the one hand, it describes a sense of awe, excitement, or reverential respect, while simultaneously, it indicates dread, shock, or the fear of a coming disturbance. Current AI hype redounds with everyday discourses, such as those that order the lived reality of everyday student tasks. It entails predictions about future impact on college education in public discourse. "In the context of generative AI, a subset of artificial intelligence, hype encompasses inflated expectations and predictions (positive and negative) about the technology's potential impact and future developments on college education" (Paulson, 2024, para. 1). The implied value system is always paradoxical, positive, and negative; however, the expectation of future impact appears to be assumed. AI hype provides for this paper a term to help shed light on a rhetorical strategy that got people to adopt generative AI, but one that also instigated a massive consumer reaction and uptake of the technology.

Researchers have studied the emergence of AI applications in education and have identified the risks. Zawacki-Richter et al. (2019) synthesized 146 articles on artificial intelligence applications in higher education, determining:

> Despite the enormous opportunities that AI might afford to support teaching and learning, new ethical implications and risks come in with the development of AI applications in higher education. For example, in times of budget cuts, it might be tempting for administrators to replace teaching by profitable automated AI solutions. Faculty members, teaching assistants, student counsellors, and administrative staff may fear that intelligent tutors, expert systems and chat bots will take their jobs. (p. 2)

Relevant is the lingering fear of AI in higher education that intensified and exhibited itself in mainstream discourse after the launch of ChatGPT.

UNESCO released the *AI and Education: Guidance for Policy Makers* report in 2021, which states that "the rapid development of Artificial Intelligence (AI) is having a major impact on education. Advances in AI-powered solutions carry enormous potential for social good and the achievement of the Sustainable Development Goals" (Miao et al., 2021, p. 4). In September 2023, UNESCO released the *Guidance for Generative AI in Education and Research* report which "looks into the possibilities for creatively using GenAI in curriculum design, teaching, learning and research activities" (Miao & Holmes, 2023, p. 7).

## Methodology

The methodological approach for this article was broadly contextual: it considered the ways education stakeholders might undergo an urgent process of cultural adaptation to generative AI, without (at the time) access to sufficient academic sources. An analysis was conducted in five main stages.

In the first stage, Part One, humanities methods were used by the author, Isabel Pedersen to address ChatGPT's appeal to students and faculty members during the early product release. Interpretative critical analysis and close reading of primary texts are used to support arguments. The interpretation describes ways that ChatGPT's alluring efficiency makes it instantly appealing. Textual content was drawn from the OpenAI website during the time of the release in 2023.

Part Two identifies news stories that sensationalized and hyped the emergence of ChatGPT at the point it was introduced to the general public. These legacy news stories impact popular understandings of the technology, including assumptions surrounding its usage in postsecondary education. As Lim (2012) notes, culture, including texts, shape our perspectives: "what one 'sees' in a text, what one regards as worth describing, and what one chooses to emphasize in a description, are all dependent on one's interpretation and understanding of the text vis-à-vis its larger social and political environment" (p. 63).

This article employs critical discourse analysis for the theoretical and methodological approach. Discourse analysis provides a means to analyze how shared values are reflected in popular media, such as news articles. Hodge and Kress (1988) assert that discourse "is the site where social forms of organization engage with systems of signs in the production of texts, thus reproducing or changing the sets of meanings and values which make up a culture" (p. 6). As a method, critical discourse analysis "relies on a collection of techniques for the study of language use as a social and cultural practice (Fairclough, 2001)" (Mullet, 2018, p.117). Critical Discourse Analysis (CDA) "like discourse analysis (DA), examines the ways in which language produces and moderates social and psychological phenomena; however, CDA emphasizes the role of language as a power resource (Willig, 2014)" (Mullet, 2018, p. 116). It "seeks to uncover, reveal, and disclose implicit or hidden power relations in discourse" (Mullet, 2018, p. 119).

In the second stage, a search was conducted by Isabel Pedersen on a corpus of mainstream legacy news sources, *The Globe and Mail* newspaper and *The Economist,* to understand how discourses frame ChatGPT and education at the time of the technology's release. This article does not imply that the legacy news sources are the only way postsecondary stakeholders get their news. These publications were chosen for their editorial codes of conduct and high standard of journalistic integrity. *The Economist* is known as a source for its data journalism; "theory, sound concepts, generally accepted indicators, indices, rankings, and so on form the basic raw material of economic journalism" (Arrese, 2022, p. 491). The ProQuest database was queried on two publications: *The Globe and Mail* newspaper and *The Economist* for the term "ChatGPT" up to May 20, 2023, with the goal to capture discourses up to, including, and shortly after the technology's launch. After the exclusion of duplicate articles by a research assistant, the set included 71 articles from *The Globe and Mail* newspaper and 20 from *The Economist*, forming a corpus of relevant articles that were saved as PDFs for analysis. While not all the news articles discussed ChatGPT or generative AI as the central topic, all mentioned it; this was manually confirmed. The point was to identify and analyze the discursive placement of an emergent technology intended for a general audience (see Table 1). An analysis was conducted in the remaining stages by two researchers (the author and a research assistant).

In the third stage, a set of search terms were used for qualitative analysis on the corpus in the broad context of education. Specifically, the following terms were searched: "education," "teach," "learn," "university," "college," "postsecondary," "researcher," "student," "courses," "class," "exam," "professor," "academic," "academia," "essay," "paper," and "lecture." This led to a smaller selection of 55 articles from the publications (11 from *The Economist* and 44 from *The Globe and Mail*). From this subset, paragraphs, phrases, and concepts were analyzed to understand the way sensationalized and exaggerated language and rhetoric—AI hype—appeared in reference to education and ChatGPT in the six months following its release.

In the fourth stage of analysis, the author identified repeated themes in the remaining subset of articles and identified four thematic categories. The subset was then reviewed by a research assistant with a critical discourse analysis approach to identify the presence of the discursive themes—declarative statements, value judgements, observations, et cetera—about AI and education. The discursive themes were qualitatively identified and counted, to ensure the study was observing actual (quantifiable) levels of prominence.

The following themes were identified in the retained articles as most prevalent: Theme One: Overwhelming Scope of Capabilities was evident in eight articles from *The Economist* and 19 from *The Globe and Mail*; Theme Two: Impending Job Loss, was identified in four articles from *The Economist* and fourteen from *The Globe and Mail;* Theme Three: Anthropomorphism of AI: was noted in one article from *The Economist* and five from *The Globe and Mail*; and Theme

Four: Vulnerability, Fear, and Existential Risk was apparent in two articles from *The Economist* and seventeen from *The Globe and Mail*.

Finally, in a fifth stage of analysis, these thematic paragraphs, phrases, and concepts were analyzed using a critical discourse analysis approach. The author selected representative quotations seen below. Then, she observed an overarching hyped rhetoric in the texts, and characterized the references to education and ChatGPT in the six months following release.

**Table 1**

*News Articles Corpus Counted According to Thematic Discourses*

|  | The Globe and Mail | The Economist |
|---|---|---|
| Corpus, Complete Set | 71 | 20 |
| Education Subset | 44 | 11 |
|  |  |  |
| Overwhelming Scope of Capabilities | 19 | 8 |
| Impending Job Loss | 14 | 4 |
| Anthropomorphism of AI | 5 | 1 |
| Vulnerability, Fear, and Existential risk | 17 | 2 |

While this article is geared to postsecondary education, the discourse analysis is designed to acknowledge the rich assemblages that educators at all levels interact with and through when adopting and adapting to educational technologies.

**Part One: Generative AI and Education: Why ChatGPT is Appealing**

Generative AI chatbots like ChatGPT are highly efficient tools for creating prose, which is one of the reasons for ChatGPT's early popularity with students, instructors, and professionals alike in postsecondary spheres (Pedersen, 2023). Chatbots can produce stylistically correct sentences, paragraphs, and documents across a multitude of genres. They can produce professional-grade visual images and video. People use them to generate programming code with natural language prompts and convert code from one programming language to another. They save students' and instructors' time. In doing so, generative AI chatbots alter the act of composition as an academic practice. Composing written answers, creating graphics, or even generating videos is automated; consequently, chatbots reduce the imaginative academic work a student needs to perform. The process is simple. One chooses a generative AI application, an AI chatbot on the Internet (e.g., ChatGPT). One prompts the chatbot in the form of a text question or an uploaded image. It answers in polite human-like terms and then returns new content.

UNESCO's *Guidance for Generative AI in Education and Research* (Miao & Holmes, 2023) unpacks how GPTs generate text in response to a prompt:

- The GPT uses statistical patterns to predict likely words or phrases that might form a coherent response to the prompt.
- The GPT identifies patterns of words and phrases that commonly co-occur in its prebuilt large data model (which comprises text scraped from the internet and elsewhere).
- Using these patterns, the GPT estimates the probability of specific words or phrases appearing in a given context.
- Beginning with a random prediction, the GPT uses these estimated probabilities to predict the next likely word or phrase in its response (p. 9).

The above quotation describes how GPTs generate prose. However, descriptions of generative AI rarely describe all the human tasks that have been eliminated. Students *no longer* perform the act of writing in this scenario. Rather than having students *thinking about the audience, searching for information, reading it, selecting and developing key points, note-taking,* and *writing* answers, generative AI chatbots reply on "prompts." Students' work is more geared to *asking* (prompting) chatbots to produce text according to a genre, *assessing* the text returned, and *copying/pasting* text answers. The final task is *structuring* a composition rather than composing it (writing it). In this formulation, students assign a responsibility to an AI agent and that agent completes a task, *answering* for the student. Students become managers of automated learning tasks. Instead of composing their own writing as discussed above, they are adapting to the idea of editing and correcting the writing (produced text) of generative AI.

OpenAI intended to develop ChatGPT as an AI assistant for educators in 2022. At the time of the ChatGPT launch, teachers of all levels were encouraged to use it despite a prominent backlash against it due to its role in facilitating student academic misconduct. In May of 2023, the OpenAI website included text on why educators might want to use it under the title, "Some examples of how we've seen educators exploring how to teach and learn with tools like ChatGPT." The website went on to list them:

1. Drafting and brainstorming for lesson plans and other activities
2. Help with design of quiz questions or other exercises
3. Experimenting with custom tutoring tools
4. Customizing materials for different preferences (simplifying language, adjusting to different reading levels, creating tailored activities for different interests)
5. Providing grammatical or structural feedback on portions of writing
6. Use in upskilling activities in areas like writing and coding (debugging code, revising writing, asking for explanations)
7. Critique AI generated text. (*Educator FAQ*, 2023, para. 2)

While the text and link are different now, OpenAI initially revealed its intent for education as a target domain and ChatGPT as a key EdTech tool. Assigning functional, professional roles for ChatGPT, OpenAI encouraged educators to use it for "brainstorming," "experimenting," and "critiquing" as well as helping students with grammar and writing. Postsecondary educators were implicated in the discourse surrounding the release as a way to make ChatGPT appear useful and credible.

From November 2022 to May 2023, during the ChatGPT release, an open education community promoted using AI in courses. The *Creating a Collection of Creative Ideas to use AI in Education* resource (Nerantzi et al., 2023) is hosted by #creativeHE. It is a community that identifies "their [productive] interactions [that] create a wide range of resources, opportunities for discussions and debates through a series of activities, including open courses and online discussions, local meetups when and where possible" (About #creativeHE, 2022, para. 4). Inviting postsecondary students and teachers to explore the embryonic stage of generative AI will help society adapt to this technology in meaningful ways (Mahmud, 2024). Another source responding to ChatGPT's release (Jones, 2023), expresses how "the advent of generative AI has ushered every academic into a new era of higher education that presents a myriad of opportunities to embrace new and powerful tools in today's learning processes" (Jones, 2023), p. 119).

The next part analyzes hyped discourses during the first phases of ChatGPT's public emergence.

**Part Two: AI Hype in Education: Why ChatGPT is Threatening**

Described above, AI hype is a rhetorical term that can be used to characterize sensationalized or hyperbolic descriptions or representations of artificial intelligence in discourses. Critical discourse analysis reveals ways that power imbalances are created through discursive manipulation. This article interprets how the ChatGPT launch delegitimizes education and educators in highly sensationalized language. The four themes identified through CDA are used as headings and include representative quotations from the education subset from *The Globe and Mail* and *The Economist*. While some of the quotations speak directly to postsecondary concerns, other aspects of education or knowledge workers, some represent AI hype as a mainstream phenomenon to provide the proper context for this paper.

*Overwhelming Scope of Capabilities*

Surprise over the scope of ChatGPT's capabilities is clearly present in the discourse:

> Already, another revolution is rushing at us: artificial intelligence. This could–and I'm not exaggerating–save our planet. (Rachman, 2022, p. A11)

> OpenAI's ChatGPT, an advanced chatbot, amassed over 100 million monthly active users and exhibited unprecedented capabilities, from crafting essays and fiction to designing websites and writing code. You'd be forgiven for thinking there's little it can't do. (Cukier, 2023, para. 3)

> A decade ago, the idea that algorithms could supplant screenwriters, displace software developers or diagnose diseases was science fiction. Now AI tools are becoming so ubiquitous that high school students are using them to cheat on tests. They're secretive, private and contain proprietary information that makes it almost impossible to learn more about their inner workings, even when they get things wrong. (Small, 2023, p. B1)

Speed of adoption, extent of capabilities, and opaque "inner workings" (Small, 2023, p. B1) formed a distinct pattern identified in the discourse, rather than critical assessment of ChatGPT. The tongue-in-cheek commentary–"You'd be forgiven for thinking there's little it can't do" (Cukier, 2023, para. 3) also betrays discomfort with ChatGPT's adoption, and fear of the unknown. Many news stories mentioned ChatGPT's ability to produce writing for cheating on assignments. Specifically, fifty-three variations of the word "cheat" appeared in *The Globe and Mail* article education subset: there were eight instances of "cheat," plus thirty-eight of "cheating," four of "cheaters," two of "cheats," and one of "cheater."

*Impending Job Loss*

The release of ChatGPT channelled the narrative that AI causes job loss, a potential harmful outcome for educators and students:

> There is a real, visceral fear among workers that their jobs could be automated away, including those of knowledge workers who once thought themselves immune. ("CROSSOVER," 2023, p. A10)

> Fourteen of the top 20 occupations most exposed to AI are teachers (foreign-language ones are near the top; geographers are in a slightly stronger position). But only the bravest government would replace teachers with AI. Imagine the headlines. ("Your job is (probably) safe from artificial intelligence," 2023, para. 20)

> After decades of blue-collar jobs being snatched up by machines, advanced chatbots are now breathing down white collars...Most exposed are industries which rely on programming and writing skills, such as legal and financial services. Teachers, especially those of languages, literature and history, are next on the list. ("ChatGPT could replace telemarketers, teachers and traders," 2023, para. 1)

The discourse foregrounds how teachers are at risk of job loss. It was previously believed that *some* jobs were at risk, now the threat is that AI teachers could replace all human teachers. The question over which government authority will counter this trend is also present, suggesting that AI operates without oversight.

*Anthropomorphism of AI*

In the discourse, ChatGPT's functions are flagged as exemplary and anthropomorphic. *Anthropomorphism* means the attribution of human traits, emotions, or intentions to non-human entities (Oxford English Dictionary, 2023):

> And then ChatGPT happened. It is now advanced enough to, for example, score in the 90th percentile on the American uniform bar exam. It can also do something even more difficult, more human –and more dangerous. It can lie. (Keller, 2023, p. B4)

> Unchecked, exponential growth in computer intelligence is in an entirely new and different category. We are on the way to building something much, much smarter than ourselves, not in the narrow sense of a machine that can follow our instructions, rapidly but moronically, but of a machine that does not need us–that can improvise, learn, adapt, even write new code for itself: a recursive, recombinant, self-contained loop of ever-expanding, ever accelerating capacity. (Coyne, 2023, p. O2)

Anthropomorphizing generative AI tasks is problematic; they don't "lie" as the above article claims, nor display any heightened cognitive ability (e.g., "smarter than ourselves"), they simply provide efficient outputs or they malfunction ("lying").

### *Vulnerability, Fear, and Existential Risk*

Several controversies concerning generative AI mark the discourse. On March 22, 2023, The Future of Life Institute asked AI experts to sign the Pause Giant AI Experiments: An Open Letter, which states: "We call on all AI labs to immediately pause for at least 6 months the training of AI systems more powerful than GPT-4" (Pause, 2023, para. 3). While these moments are not discussed in detail in this article, the central point is to highlight the way fears were instantiated in the discourse following ChatGPT's launch. Beliefs, including the future sentience of AI, the end of human civilization, and an abstract fear over the danger of AI, appeared at certain intervals. Figure 1 displays news events that helped instigate hyped language, such as these quotations:

> Over the past couple of years new AI tools have emerged that threaten the survival of human civilisation. (Harari, 2023, para. 1)

> When it comes to artificial intelligence, however, we are confronted with an altogether different challenge. It is not our intelligence that is being expanded, but our creation's– the computer–to the point that some have begun to fear it will surpass or even replace us. (Coyne, 2023, p. O2)

> 'At the absolute frontier, where we're creating systems that we don't yet know exactly what the emerging capabilities will be, I think there should be some sort of global licensing and regulatory framework, in the same way we do for other superdangerous, superhigh-potential technology,' Mr. Altman said. (Castaldo, 2023, p. A4)

Caught up in this chaos, target domains such as education, healthcare, and the arts are relegated as agentless. Inventors and CEOs do not trust their own AI products not to harm society. Strategically, these activities are likely to further corporate tactics to gain a competitive advantage and profit.

**Figure 1**

*Dates and Descriptions of Sensational International News Events Leading to AI hype*

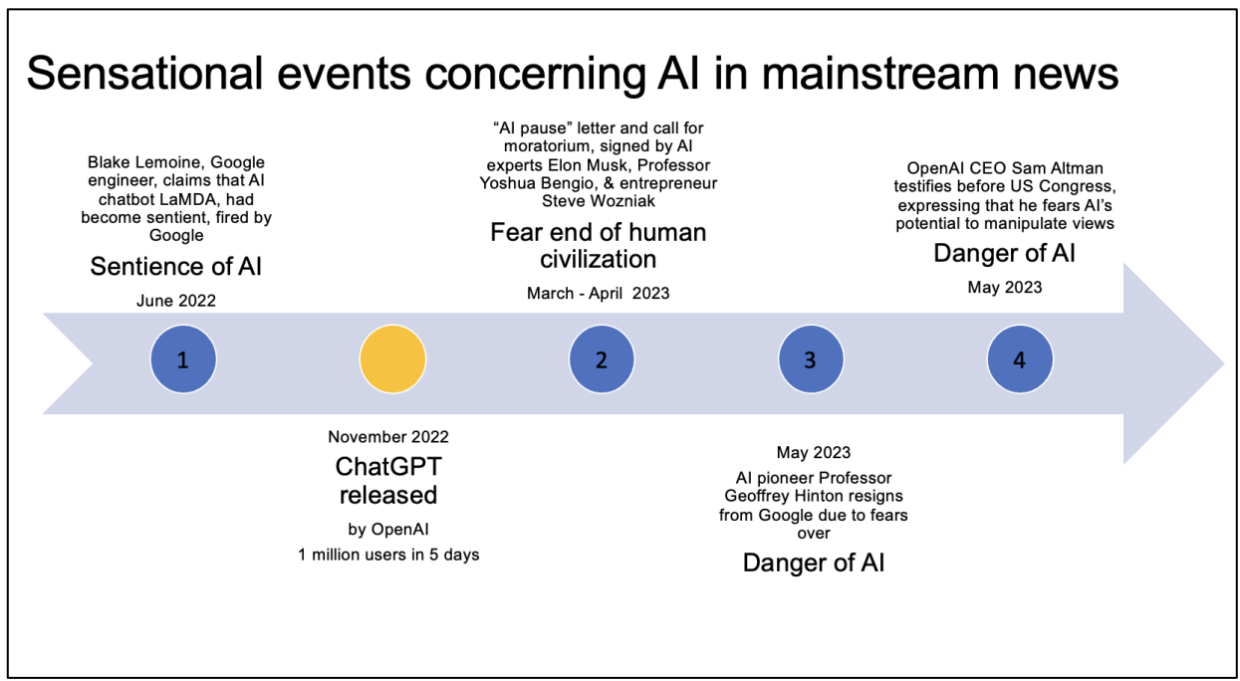

## Discussion

News surrounding OpenAI's launch of ChatGPT prompted a significant techno-cultural phenomenon. Rather than undertaking a measured approach and adaptation to a suite of educational tools, postsecondary educators were met with a panicked discourse circulating in traditional news outlets surrounding generative AI. The first theme, *Overwhelming Scope of Capabilities*, was the most dominant pattern in the discourse. Journalists often used frank terms about ChatGPT's ability to help students cheat. For example, Ben Harvey, a Toronto-based librarian writes in *The Globe and Mail* on December 15, 2022:

> How should we think, then, about this latest compelling development in AI? Certainly, professors and teachers are wise to worry, because this software appears to be the perfect essay mill, as it allows users to slap together passable critical writing in seconds. (p. A11).

The news story appears less than three weeks after the ChatGPT release, demonstrating harsh commentary on the scope of its capabilities.

A finding of the critical discourse analysis was that AI hype is amplified by overlapping concerns. This article identified four themes as if they are separate; however, they function intertextually. For example, as the *Overwhelming Scope of Capabilities* theme is sensationalized, professionals are framed as weakened. As a consequence, the theme of *Impending Job Loss* for teachers appears as a seemingly logical next step, e.g., "jobs being snatched up by machines" ("ChatGPT could replace telemarketers, teachers and traders," 2023, para. 1). Much more broadly, the *Anthropomorphism of AI* theme helps to buttress the news events that lead to the *Vulnerability, Fear, and Existential Risk of AI* theme, e.g., "We are on the way to building something much, much smarter than ourselves" (Coyne, 2023, p. O2). Comparing students' intelligence to machine intelligence does not make sense. Like any machine, a computer is fundamentally designed to make human work more efficient. In education, making assumptions that a technology behaves as if it has intentions is deceptive and persuasive, leading to misinterpretations of students' usage of the technology and their abilities. Adapting to the dynamic nature of the tools and the EdTech market will be an evolving task for postsecondary institutions.

At the same time, anthropomorphizing AI is a staple of hyped AI discourse since chess grandmaster Gary Kasparov beat and then lost to IBM's Deep Blue in 1997. In his 2017 book, *Deep Thinking*, Kasparov makes this point clear. He writes,

> This sort of thinking is a trap into which every generation falls when it comes to machine intelligence. We confuse performance—the ability of a machine to replicate or surpass the results of a human—with method, how those results are achieved. (p. 26).

He adds, "this romanticizing and anthropomorphizing of machine intelligence is natural" emphasizing that it is simply a means to explain the technology (p. 26). In *Atlas of AI* (2021), Kate Crawford expresses that the "mythologies" of AI anthropomorphism "are particularly strong in the field of artificial intelligence, where the belief that human intelligence can be formalized and reproduced by machines has been axiomatic since the mid-twentieth century" (p. 13). She iterates that AI "is not an objective, universal, or neutral computational technique.... Its systems are embedded in social, political, cultural, and economic worlds, shaped by humans, institutions, and imperatives that determine what they do and how they do it" (p. 220).

Going forward, generative AI tools ought to be developed for education according to the needs of students and instructors and not the fear of the unknown. George Veletsianos points out that "ChatGPT is not a threat, but an opportunity for the education system to renew itself, to imagine a better world for its students" (Veletsianos, 2023a, p. A11). UNESCO's *Recommendation on the Ethics of Artificial Intelligence* (2021) advises that Member States "encourage research initiatives on the responsible and ethical use of AI technologies in teaching, teacher training and e-learning" (p. 34.). However, it also discusses limitations on AI systems in learning so that students are protected; for example, they "should be subject to strict requirements when it comes to the monitoring, assessment of abilities, or prediction of the learners' behaviours" (UNESCO, 2021, p. 34).

## Conclusion

Concentrating on ChatGPT's release, this article argues that AI hype has altered the means through which postsecondary education is adopting generative AI. It includes the analysis of 55 articles from *The Economist* and *The Globe and Mail,* a corpus of texts that place education within a broader context of mainstream news. AI hype is observed through two seemingly oppositional reasons. On the one hand, generative AI is a highly efficient tool that replaces traditional and foundational work for students, instructors and education professionals alike. On the other hand, participation, adoption, and planning for its use is mired in panic for the future of education. The qualitative discourse analysis provides means to unpack four sensational themes.

Next steps for research in postsecondary spheres ought to include more attention to the expansion of emergent generative AI technologies for education that will be released. In January 2023, OpenAI announced the availability of

> developer APIs for ChatGPT and [other] AI models that will let developers integrate them into their apps. An API (Application Programming Interface) is a set of protocols that allows different computer programs to communicate with each other. In this case, app developers can extend their apps' abilities with OpenAI technology. (Edwards, 2023, para. 1).

The reality is that generative AI is rapidly being integrated into foundational components of the Internet to ensure connectivity, and ultimately facilitate further efficiency. Including students, communities, and public rights holders in educational tools discovery will better guide policymaking.

Higher education ought to be planning for the next ten to twenty years of technological emergence that will involve ongoing adaptation to AI writing in classrooms, research fora, and in teaching and learning professional practices. Good sources for teachers seeking novel ideas for curriculum include the Open Educational Resources (OER) Initiative that hosts the *Writing and Artificial Intelligence: An Open Educational Resources (OER) Guide* (Open Educational Resources Initiative, 2023), and the #creativeHE community. Higher education is undergoing a process to reimagine educational technologies as inclusive sites for collaboration, networking, co-creation, and student empowerment (Killam et al., 2023; Mahmud, 2024; Romero-Hall et al., 2023). Killam et al. (2023) defines course co-creation as "an open practice that engages learners and educators in shared decision-making about any aspect of course design" (p. 2). Educational technology like ChatGPT—proposed, marketed, promoted for education—needs to be embedded in ongoing work on digital literacies (Davis et al., 2021; De Laat et al, 2020; Tham et al., 2021). AI Literacy, critical media literacy, civic engagement, and ethically-aligned adoption and assessment of AI writing tools will be needed. Learning will evolve along with the tools we place in the hands of students, teachers and professionals.

The adoption of AI technologies also needs more input from traditionally marginalized communities. For example, drawing on the responses of students who self-identify as Indigenous, King and Brigham (2023) advise:

> Rather than expecting Indigenous students to adapt to the existing post-secondary educational structures, we argue that it is necessary for institutions to reimagine themselves.... This includes respecting and promoting Indigenous-preferred languages in universities, increasing Indigenous-informed curriculum materials and educational resources, and making Indigenous perspectives and knowledges visible throughout university campuses. (p. 52)

King and Brigham foreground a few co-design values for marginalized communities that should be upheld as we adopt generative AI. The adoption of educational technologies cannot proceed as a reaction to mainstream AI hype. Postsecondary institutions can instead draw on valuable extant research on emergent technologies and inclusive teaching and learning, as they reimagine its uses.

Finally, researchers, rightsholders, and community members need to be given space to weigh in on learning outcomes, course development, assessment strategies, course and program policies, and even Learning Management System (LMS) development, to ensure marginalized communities are included in the adoption and adaptation to generative AI and other large language model technologies.

## Notes

1. Generative AI writing is vulnerable to producing biased, discriminatory, and racist results, as well as online hate, in produced text, images, or video due to training sources. It is often factually inaccurate and can promote false information or fictionalize an academic source.
2. This paper does not attend to the issue that certain generative AI companies might be breaking copyright or intellectual property laws due to the training of LLMs.
3. Most institutions have produced course policies to assist instructors on assessing academic misconduct in postsecondary institutions concerning generative AI.

## Open Researcher and Contributor Identifier (ORCID)

Isabel Pedersen  https://orcid.org/0000-0002-9317-0812

## Acknowledgements

The author would like to express her appreciation of the diligent work conducted by senior research assistant, Dr. Kristen Aspevig on this case study. She thanks the reviewers and editors for their constructive feedback, insightful comments, and careful editing of the manuscript.

## Ethics Statement

An ethics review was not applicable because there was no research involving human subjects.

## Conflict of Interest

The author does not declare any conflict of interest.

## Data Availability Statement

Publications were downloaded using ProQuest database university searches.

## References

*About creativeHE.* (2022). #creativeHE. Retrieved Nov. 30, 2023, from https://creativehecommunity.wordpress.com/about/

Akbari, N. (2023, February 28). The AI cheating crisis: Education needs its anti-doping movement. *EducationWeek.* https://www.edweek.org/technology/opinion-the-ai-cheating-crisis-education-needs-its-anti-doping-movement/2024/02

Arrese, A. (2022). "In the beginning were the data": Economic journalism as/and data journalism. *Journalism Studies. 23*(4), 487–505. https://doi.org/10.1080/1461670X.2022.2032803

Castaldo, J. (2023, May 16). OpenAI CEO says global regulation needed to mitigate potential AI harms. *The Globe and Mail*, p. A4. https://www.theglobeandmail.com/business/article-openai-chatgpt-ceo-altman-speech/

ChatGPT could replace telemarketers, teachers and traders: Here's why that is no bad thing. (2023, Apr 14). *The Economist*. https://www.economist.com/graphic-detail/2023/04/14/chatgpt-could-replace-telemarketers-teachers-and-traders

Chami, G. (2023, Oct 23) Artificial intelligence and academic integrity: Striking a balance. *Times Higher Education.* Retrieved March 17, 2023, from https://www.timeshighereducation.com/campus/artificial-intelligence-and-academic-integrity-striking-balance#

Coyne, A. (2023, May 6). Dr. Frankenstein awakens to the AI monster he has made: The exponential growth of artificial intelligence could have catastrophic consequences, and leading researchers in the field are speaking out about their fears. but it's not too late to change course. *The Globe and Mail*, p. O2. https://www.theglobeandmail.com/opinion/article-dr-frankenstein-awakens-to-the-ai-monster-he-has-made/

Crawford, K. (2021). *Atlas of AI: Power, politics and the planetary costs of artificial intelligence*. Yale University Press.

CROSSOVER: The robots are coming. Is Ottawa ready? (2023, April 5). *The Globe and Mail*, p. A10. Retrieved Nov. 26, 2023, from https://www.theglobeandmail.com/opinion/editorials/article-the-robots-are-coming-is-ottawa-ready/

Cukier, K. (2023, April 20). Generative artificial intelligence on our "Babbage" podcast: Essential listens. *The Economist*. https://www.economist.com/AI-pods

Davis, K., Stambler, D., Veeramoothoo, C., Ranade, N., Hocutt, D., Tham, J., Misak, J., Duin, A. H., & Pedersen, I. (2021). Fostering student digital literacy through the fabric of digital life. *Journal of Interactive Technology and Pedagogy*. https://jitp.commons.gc.cuny.edu/fostering-student-digital-literacy-through-the-fabric-of-digital-life/

de Laat, M., Joksimovic, S., & Ifenthaler, D. (2020). Artificial intelligence, real-time feedback and workplace learning analytics to support in situ complex problem-solving: A commentary. *The International Journal of Information and Learning Technology*, *37*(5), 267–277. https://doi.org/10.1108/IJILT-03-2020-0026

Duin, A. H., & Pedersen, I. (2021). *Writing futures: Collaborative, algorithmic, autonomous.* Springer-Verlag. https://doi.org/10.1007/978-3-030-70928-0

Duin, A. H., & Pedersen, I. (2023). *Augmentation technologies and artificial intelligence in technical communication: Designing ethical futures.* Routledge, Taylor & Francis.

*Economist Impact: Working together with our partners for progress.* (2022). https://impact.economist.com/

*Educator FAQ.* (2023) Open AI. Retrieved May 2023, from https://platform.openai.com/docs/chatgpt-education

Edwards, B. (2023, Mar 1.) *AI For hire — ChatGPT and whisper APIs debut, allowing*

*devs to integrate them into apps.* Ars Technica. https://arstechnica.com/information-technology/2023/03/chatgpt-and-whisper-apis-debut-allowing-devs-to-integrate-them-into-apps/

Field, H. (2023, November 6). *Microsoft-backed OpenAI announces GPT-4 Turbo, its most powerful AI yet.* CNBC. https://www.cnbc.com/2023/11/06/openai-announces-more-powerful-gpt-4-turbo-and-cuts-prices.html

Floridi, L., & Chiriatti, M. (2020). GPT-3: Its nature, scope, limits, and consequences. *Minds & Machines 30*, 681–694. https://doi.org/10.1007/s11023-020-09548-1

Goyal, M., Varshney, S., & Rozsa, E. (2023). *What is generative AI, what are foundation models, and why do they matter?* IBM Think. https://www.ibm.com/blog/what-is-generative-ai-what-are-foundation-models-and-why-do-they-matter/

Harari, Y. (2023, April 28). Noah Harari argues that AI has hacked the operating system of human civilisation: Artificial intelligence. *The Economist*. https://www.economist.com/by-invitation/2023/04/28/yuval-noah-harari-argues-that-ai-has-hacked-the-operating-system-of-human-civilisation

Harvey, B. (2022, Dec. 13). Despite the rise of AI, we cannot always outsource and optimize thinking. *The Globe and Mail*, p. A11.

Hodge, R., & Gunther K. (1988). *Social semiotics.* Cornell University Press.

Intelligent. (2023). One-third of college students used ChatGPT for schoolwork during the 2022-23 academic year. https://www.Intelligent.Com/One-Third-Of-College-Students-Used-Chatgpt-For-Schoolwork-During-The-2022-23-Academic-Year/

Jones, M. (2023). Preserving academic integrity in the age of artificial intelligence: Redesigning courses to combat AI-assisted plagiarism. *International Dialogues on Education Journal*, *10*(1), 101–123. https://www.idejournal.org/index.php/ide/article/view/296

Kasparov, G. (2017). *Deep thinking: Where machine intelligence ends and human creativity begins.* PublicAffairs.

Keller, T. (2023, May 5). Artificial intelligence is coming, and it may be after more than just your job. *The Globe and Mail*, p. B4. https://www.theglobeandmail.com/business/commentary/article-artificial-intelligence-chatgpt-dangers/

Killam, L., Chumbley, L., Kohonen, S., Stauffer, J., & Mitchell, J. (2023). Co-creation during a course: A critical reflection on opportunities for co-learning. *The Open/Technology in Education, Society, and Scholarship Association Journal*, *3*(1), 1–12. https://doi.org/10.18357/otessaj.2023.3.1.43

King, A., & Brigham, S. (2023). "I was like an alien": Exploring how Indigenous students succeed in university studies. *Canadian Journal of Higher Education / Revue canadienne d'enseignement supérieur*, *52*(4), 41–55. https://doi.org/10.47678/cjhe.v52i4.189757

Larsen, B. & Narayan, J. (2023, Jan. 9). *Generative AI: A game-changer societyneeds to be ready for.* World Economic Forum. Retrieved April 5, 2023, from https://www.weforum.org/agenda/2023/01/davos23-generative-ai-a-game-changer-industries-and-society-code-developers/

Lim, L. (2012). Ideology, rationality and reproduction in education: A critical discourse analysis. *Discourse: Studies in the Cultural Politics of Education*, *35*, 61–76. https://doi.org/10.1080/01596306.2012.739467

Mahmud, S. (Ed.). (2024). Academic integrity in the age of artificial intelligence. *IGI Global.* https://doi.org/10.4018/979-8-3693-0240-8

Malik, A. (2023, November 6). *OpenAI's ChatGPT now has 100 million weekly active users.* TechCrunch. https://techcrunch.com/2023/11/06/openais-chatgpt-now-has-100-million-weekly-active-users

McKee, H., & Porter, J. (2021). Intertext—Writing machines and rhetoric. In A. H. Duin & I. Pedersen (Eds.), *Writing futures: Collaborative, algorithmic, autonomous* (pp. 47–52). Springer.

Miao, F., & Holmes, W. (2023). *Guidance for generative AI in education and research.* United Nations Educational, Scientific and Cultural Organization (UNESCO). https://doi.org/10.54675/EWZM9535

Miao, F., Holmes, W., Ronghuai, H., & Hui, Z. (2021). *AI and education: guidance for policy-makers.* UNESCO. https://doi.org/10.54675/PCSP7350

Mullet, D. (2018). A general critical discourse analysis framework for educational research. *Journal of Advanced Academics*, *29*(2), 116–142. https://doi.org/10.1177/1932202X18758260

Nerantzi, C., Abegglen, S., Karatsiori, M., & Martínez-Arboleda, A. (Eds.). (2023). *101 creative ideas to use AI in education, A crowdsourced collection.* #creativeHE. https://creativehecommunity.wordpress.com/2023/06/23/oa-book-101-creative-ideas-to-use-ai-in-education/

Omrow, D. A. (2018). It is not easy being green: A critical discourse and frame analysis of environmental advocacy on American television. *Journal of Media and Communication Studies*, *10*(3), 14–24. https://doi.org/10.5897/JMCS2018.0609

Open Educational Resources Initiative. (2023, November). *Writing and artificial intelligence: An open educational resources (OER) guide.* ASCCC Open Educational Resources Initiative, Academic Senate for California Community Colleges. https://asccc-oeri.org/2023/11/28/writing-and-artificial-intelligence-an-open-educational-resources-oer-guide/

Oxford English Dictionary. (2023). Anthropomorphism. In *Oxford English Dictionary* online, Oxford University Press. https://www.oed.com/search/dictionary/?scope=Entries&q=anthropomorphism

Paulson, E. (2024, February 22). Decoding AI hype for teaching and learning. Conestoga. https://tlconestoga.ca/decoding-ai-hype-for-teaching-and-learning/

*Pause giant AI experiments: An open letter.* (2023, March 22). Future of Life Institute. Retrieved Dec. 21, 2023, from https://futureoflife.org/open-letter/pause-giant-ai-experiments/

Pedersen, I. (2023). The rise of generative AI and enculturating AI writing in postsecondary education. *Frontiers in Artificial Intelligence*, *6*. https://doi.org/10.3389/frai.2023.1259407

Rachman, T. (2022, December 28). Artificial intelligence has already shifted the balance of power – The question now is, can humanity catch? *The Globe and Mail,* p. A11. https://www.theglobeandmail.com/opinion/article-artificial-intelligence-has-already-shifted-the-balance-of-power-will/

Romero-Hall, E., Gomez-Vasquez, L., Forstmane, L., Ripine, C., & Dias da Silva, C.

(2023). The complexities of using digital social networks in teaching and learning. *The Open/Technology in Education, Society, and Scholarship Association Journal*, *3*(1), 1–18. https://doi.org/10.18357/otessaj.2023.3.1.48

Small, T. (2023, May 12). The downside of AI: Former Google scientist Timnit Gebru warns of the technology's built-in biases. *The Globe and Mail*, p. B1. https://www.theglobeandmail.com/business/article-the-downside-of-ai-former-google-scientist-timnit-gebru-warns-of-the/

Tham, J. C. K., Burnham, K.D., Hocutt, D.L., Ranade, N., Misak, J., Duin, A.H., Pedersen, I, & Campbell, J. L. (2021). Metaphors, mental models, and multiplicity: Understanding student perception of digital Literacy. *Computers and Composition, 59*, Article 102628. https://doi.org/10.1016/j.compcom.2021.102628

*The Economist media kit* (2023). Improvado. Retrieved December 18, 2023, from https://improvado.io/resources/economist-media-kit

*The Globe and Mail newspaper media kit* (2023). Globe Media Group. Retrieved December 18. 2023 from https://globelink.ca/wp-content/uploads/2023/01/The-Globe-and-Mail-Newspaper-MediaKit-2023.pdf

Thormundsson, B. (2023, Apr 25). *ChatGPT - Statistics & facts*. Statista.

United Nations Educational, Scientific and Cultural Organization (2021). *Recommendation on the ethics of artificial intelligence.* UNESCO. Retrieved Dec. 20, 2023 from https://unesdoc.unesco.org/ark:/48223/pf0000381137

Veletsianos, G. (2023a, April 25). Making ChatGPT detectors part of our education system puts surveillance over trust. *The Globe and Mail*, p. A11. https://www.theglobeandmail.com/opinion/article-making-chatgpt-detectors-part-of-our-education-system-prioritizes/

Veletsianos, G. (2023b). *Generative artificial intelligence in Canadian post-secondary education: AI policies, possibilities, realities, and futures, 2023 Special Topics Report*. Canadian Digital Learning Research Association. https://www.d2l.com/resources/assets/cdlra-2023-ai-report/

Your job is (probably) safe from artificial intelligence: Why predictions of an imminent economic revolution are overstated. (2023, May 7). *The Economist*. https://www.economist.com/finance-and-economics/2023/05/07/your-job-is-probably-safe-from-artificial-intelligence

Zawacki-Richter, O., Marín, V.I., Bond, M., & Gourverneur, F. (2019). Systematic review of research on artificial intelligence applications in higher education – Where are the educators? *International Journal of Educational Technology in Higher Education*, *16*(39), Article 39.  https://doi.org/10.1186/s41239-019-0171-0